\documentclass[prb,twocolumn,preprintnumbers,amsmath,amssymb,showpacs,
nofootinbib,floatfix]{revtex4}

\usepackage{graphicx,bm}

\makeatletter
\def\graphicscale{\twocolumn@sw{0.3}{0.4}}
\def\graphicthreescale{\twocolumn@sw{0.3}{0.4}}

\begin{document}

\title{Critical parameters from trap-size scaling in trapped particle systems}

\author{Giacomo Ceccarelli,$^{1}$ 
Christian Torrero,$^2$ and Ettore Vicari$^1$ }

\affiliation{$^1$Dip. di Fisica dell'Universit\`a di Pisa and
INFN, Largo Pontecorvo 2, I-56127 Pisa,
Italy}

\affiliation{$^2$Dip. di Fisica dell'Universit\`a di Parma and INFN,
viale Usberti 7/A, I-43124 Parma, Italy
}

\date{January 1, 2013}

\begin{abstract}
We investigate the critical behavior of trapped particle systems at
the low-temperature superfluid transition. In particular, we consider
the three-dimensional Bose-Hubbard model in the presence of a trapping
harmonic potential coupled with the particle density, which is a
realistic model of cold bosonic atoms in optical lattices.  We present
a numerical study based on quantum Monte Carlo simulations, analyzed
in the framework of the trap-size scaling (TSS).

We show how the critical parameters can be derived from the trap-size
dependences of appropriate observables, matching them with TSS.  This
provides a systematic scheme which is supposed to exactly converge to
the critical parameters of the transition in the large trap-size
limit.  Our numerical analysis may provide a guide for experimental
investigations of trapped systems at finite-temperature and quantum
transitions, showing how critical parameters may be determined by
looking at the scaling of the critical modes with respect to the trap
size, i.e.  by matching the trap-size dependence of the experimental
data with the expected TSS Ansatz.

\end{abstract}


\maketitle



\section{Introduction}

The theory of critical phenomena~\cite{Wilson-82,ZJ-book} at phase
transitions generally applies to homogenous systems.  However,
homogenous conditions are often an ideal limit of experimental
conditions.  Thus, the study of the effects of inhomogenous features
is often essential to achieve a correct interpretation of the
experimental data at the transition between different phases, in order
to obtain reliable estimates of the critical parameters, such as the
critical temperature, universal critical exponents, etc.

In this paper we focus on quantum systems of interacting particles in
the presence of an external space-dependent potential coupled to the
particle density, which effectively traps the particles within a
limited region of space.  The presence of a harmonic trap is a common
feature of the experimental realizations of the Bose-Einstein
condensation (BEC) in diluted atomic vapors~\cite{CWK-02} and
experiments of cold atoms in optical lattices created by laser-induced
standing waves~\cite{BDZ-08,GPS-08}, which have provided a great
opportunity to investigate the interplay between quantum and
statistical effects in particle systems.

The critical behavior arising from the formation of BEC has been
investigated experimentally in a trapped atomic
system~\cite{DRBOKS-07}, observing an increasing correlation length
compatible with what is expected at a continuous transition.  However,
the inhomogeneity due to the trapping potential drastically changes,
even qualitatively, the general features of the behavior at a phase
transition.  For example, the correlation functions of the critical
modes do not develop a diverging length scale in a trap.
Nevertheless, even in the presence of the trap, and in particular when
the trap gets large, we may still observe a critical regime, although
distorted. In experiments of trapped
particle systems aimed to investigate their many-body critical
behaviors at quantum and finite-temperature phase transitions, an
accurate determination of the critical parameters, such as the
critical temperature, critical exponents, etc..., calls for a
quantitative analysis of the trap effects.  This issue has been much
discussed within theoretical and experimental investigations, see e.g.
Refs.~\onlinecite{DSMS-96,WATB-04,RM-04,FWMGB-06,NCK-06,DZZH-07,HK-08,
GZHC-09,BB-09,Taylor-09,CV-09,BSB-09,ZKKT-09,RBTS-09,HR-10,
Trotzky-etal-10,CV-10,ZH-10,HZ-10,PPS-10,PPS-10b,NNCS-10,
ZKKT-10,CV-10b,QSS-10,FCMCW-11,ZHTGC-11,CV-11,MDKKST-11,HM-11,CTV-12,
Pollet-12,CT-12,KLS-12,CR-12}.

Around the transition point, the critical behavior 
is expected to show a power-law scaling with respect to the trap size,
which we call trap-size scaling~\cite{CV-09,CV-10} (TSS), controlled
by the universality class of the phase transition of the homogenous
system.  TSS has some analogies with the standard finite-size scaling
(FSS) theory for homogenous systems~\cite{FBJ-73,Cardy-88,GKMD-08},
with two main differences: the inhomogeneity due to the
space-dependence of the external field, and a nontrivial power-law
dependence of the correlation length $\xi$ when increasing the trap
size $l$ at the critical point, i.e. $\xi\sim l^\theta$ where $\theta$
is the universal {\em trap} exponent.

In this paper we show how TSS can be exploited to determine the
critical parameters, by analyzing the
trap-size dependence of observables related to the critical modes
around the center of the trap.  The main advantage of this approach is
that it is supposed to be exact in the large trap-size limit, thus
providing a systematic scheme to control and improve the accuracy of
the results, without appealing to further assumptions and
approximations, such as mean-field and local-density approximations.
As we shall see, the method resembles standard FSS techniques, which
are routinely used to obtain accurate estimates of the critical
parameters in homogenous systems, by looking at the asymptotic
scaling behavior with respect to the size of the system, see, e.g.,
Refs.~\onlinecite{PV-02,Hasenbusch-01}.

We investigate this issue at the finite-temperature superfluid
transition of the three-dimensional (3D) Bose-Hubbard (BH)
model~\cite{FWGF-89}, which is particularly relevant for cold-atom
experiments because it describes bosonic atoms in optical
lattices~\cite{JBCGZ-98}.  For this purpose we present a numerical
analysis based on quantum Monte Carlo (QMC) simulations of the 3D BH
model with an external harmonic potential coupled to the particle
density.  

This study may be useful for experimental investigations
at finite-temperature and quantum transitions, suggesting some
effective recipes to determine the critical parameters from the
scaling of the observables related to the critical modes with respect
to the trap size, i.e.  by matching the trap-size dependence of the
experimental data with the expected TSS Ansatz, similarly to
experiments probing FSS behavior in $^4$He at the superfluid
transition~\cite{GKMD-08}.

The paper is organized as follows.  In Sec.~\ref{tssgen} we present
the general features of the TSS in trapped bosonic particle systems at
the finite-temperature superfluid transition driven by the formation
of a BEC, such as those described by the 3D BH model.
Sec.~\ref{cpfss} presents a numerical study of the superfluid
transition of the 3D BH model in the hard-core limit, using standard
FSS techniques, which allows us to verify the 3D XY universality class
of the critical behavior, and to accurately determine the critical
point.  In Sec.~\ref{critss} we show how the critical parameters can
be estimated by a TSS analysis of numerical QMC data of the trapped BH
model. Finally, in Sec.~\ref{conclusions} we draw our conclusions.
App.~\ref{thetaest} reports some details of the computation of the
trap exponent $\theta$ at the 3D superfluid transition of bosonic
systems.

\section{TSS at the superfluid transition of the 3D Bose-Hubbard model}
\label{tssgen}

\subsection{The phase diagram  of the 3D BH  model}
\label{stbh}

\begin{figure}[tbp]
\includegraphics*[scale=\graphicscale]{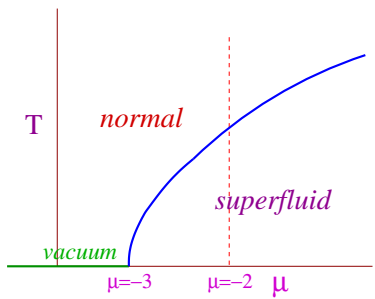}
\caption{(Color online) Sketch of the $T$-$\mu$ phase diagram of the
3D BH model (\ref{bhm}) in the hard-core $U\to\infty$ limit. The
superfluid transition line starts at $(T=0,\mu=-3)$, which is the
location of a quantum transition from the vacuum state to the
superfluid phase, and ends at another quantum transition at
$(T=0,\mu=3)$ (not shown in the figure) between the superfluid and
Mott phases.  The dashed line at $\mu=-2$ shows the line along which
we numerically investigate the critical behavior.
}
 \label{phdia}
\end{figure}

Systems of bosonic atoms in optical lattices can be modeled by the 3D
BH model~\cite{JBCGZ-98,BDZ-08}, defined by the
Hamiltonian~\cite{FWGF-89}
\begin{eqnarray}
H_{\rm BH}  &=& - {J\over 2} \sum_{\langle ij\rangle} (b_i^\dagger b_j+
b_j^\dagger b_i) + \label{bhm}\\
&+&{U\over 2} \sum_i n_i(n_i-1) - \mu \sum_i n_i\,,
\nonumber
\end{eqnarray}
where $b_i$ is the bosonic operator, $n_i\equiv b_i^\dagger b_i$ is
the particle density operator, and the sums run over the bonds
${\langle ij \rangle }$ and the sites $i$ of a cubic lattice.  The
basic observables are related to the particle density
\begin{equation}
\rho({\bf x}) \equiv \langle n_{\bf x} \rangle,\label{rhodef}
\end{equation}
and the correlation functions of the bosonic field and the particle
density,
\begin{eqnarray}
&&G_b({\bf x},{\bf y}) \equiv \langle b_{\bf x}^\dagger 
b_{\bf y} \rangle,
\label{gbdef}\\
&&G_n({\bf x},{\bf y}) \equiv 
\langle n_{\bf x} n_{\bf y} \rangle - 
\langle n_{\bf x} \rangle \langle n_{\bf y} \rangle .
\label{gndef}
\end{eqnarray}
In actual experiments the particle density and its correlations, such
as $G_n$, can be measured by the so-called {\em in situ} density image
techniques, see e.g. Refs.~\onlinecite{GZHC-09,BGPFG-09,HZHTGC-11}.
Some information on the correlation function $G_b$, and in particular
the related momentum distribution
\begin{equation}
n({\bf k}) = \sum_{{\bf x},{\bf y}}
 e^{i{\bf k}({\bf x}-{\bf y})} G_b({\bf x},{\bf y}),
 \label{nkdef}
\end{equation}
can be inferred from the
interference patterns of absorption images after a time-of-flight
period in the large-time ballistic regime, see
e.g. Ref.~\onlinecite{BDZ-08}.

The $T$-$\mu$ phase diagram of the 3D BH model presents a finite-$T$
transition line separating the normal-fluid phase and the
low-temperature superfluid phase.  The finite-$T$ superfluid
transition is characterized by the accumulation of a macroscopic
number of bosonic atoms in a single quantum state, giving rise to the
BEC.  The condensate wave function naturally
provides the complex order parameter $\psi(x)$ of the phase transition
and its relevant U(1) symmetry.  These global features characterize
the 3D XY universality class which describes the universal critical
behavior of a wide class of systems, see, e.g.,
Ref.~\onlinecite{PV-02}.  Numerical studies of the phase diagram of
the 3D BH model are reported in Refs.~\onlinecite{CPS-07,CR-12}.

In Fig.~\ref{phdia} we sketch the $T$-$\mu$ phase diagram of the 3D BH
model in the hard-core $U\to\infty$ limit, implying that the particle
number $n_i$ per site is restricted to the values $n_i=0,1$. The
finite-$T$ transition line connects two $T=0$ quantum critical points
at $\mu=\pm 3$.  The $T=0$ quantum transition at $\mu=-3$ separates
the vacuum state and the superfluid phase, while the one at $\mu=3$
separates the superfluid phase from a Mott phase. In both cases the
quantum critical behavior is essentially mean field in three spatial
dimensions~\cite{FWGF-89,Sachdev-book}.

The critical behavior described by the 3D XY universality class is
characterized by two relevant parameters $\tau$ and $h$, associated
with the temperature $T$, i.e., $\tau\sim T/T_c-1$, and the external
field $h$ coupled to the order parameter.  Their renormalization-group
(RG) dimensions, $y_\tau=1/\nu$ and $y_h=(5-\eta)/2$ respectively, are
related to the critical exponent $\nu$ of the correlation length and
to the exponent $\eta$ describing the power-law decay of the two-point
function of the order parameter at $T_c$.  The critical exponents
$\nu$ and $\eta$ are known with great accuracy from theoretical
calculations, see, e.g., the results reported in
Refs.~\onlinecite{PV-02,GZ-98}, and experiments at the $^4$He
superfluid transition~\cite{Lipa-etal-96}.  Recent theoretical
estimates of the critical exponents are~\cite{CHPV-06,BMPS-06}
\begin{equation}
\nu=0.6717(1),\qquad \eta=0.0381(2).
\label{critexpXY}
\end{equation}

\subsection{TSS in the presence of the trap}
\label{tsstrap}

A common feature of the experiments with cold atoms~\cite{BDZ-08} is
the presence of an external potential $V$ coupled to the particle
density, which traps the particles within a limited space region.  In
experiments $V$ is usually effectively harmonic.  We consider a
harmonic rotationally-invariant potential
\begin{eqnarray}
V(r)= v^2 r^2,\label{potential}
\end{eqnarray}
where $r\equiv |{\bf x}|$ is the distance from the center of the trap,
which we locate at the origin of the axis, ${\bf x}=0$.  This trapping
force gives rise to a further term in the Hamiltonian:
\begin{eqnarray}
H_{\rm tBH} = H_{\rm BH} + \sum_i V(r_i) n_i. \label{bhmt}
\end{eqnarray}
Far from the origin the potential $V(r)$ diverges, therefore $\langle
n_i\rangle$ vanishes and the particles are trapped.  
We define the trap size by
\begin{equation}
l\equiv \sqrt{J}/v. 
\label{trside}
\end{equation}
This definition naturally arises~\cite{CV-10b,BDZ-08,RM-04,DSMS-96}
 when we consider the {\em thermodynamic} limit, which is generally
 defined by the limit $N,l\to\infty$ keeping $N/l^3$ fixed ($N$ is the
 number of particles).
 The critical behaviors in the presence of the trap are generally
 studied in this limit, which  corresponds to keeping the chemical
 potential $\mu$ fixed,
 as in Eq.~(\ref{bhm}),  while increasing the trap size.
 In the following,
 we set $J=1$ so that $l=1/v$.  We consider the model at fixed
 chemical potential $\mu$, so we will skip its dependence in the
 following formulas.

In the presence of a harmonic trap, the large trap-size behavior at
the transition can be described in the framework of TSS.  The trapping
potential (\ref{potential}) coupled to the particle density, as in
Eq.~(\ref{bhmt}), significantly affects the critical modes,
introducing another length scale (\ref{trside}).  Within the TSS
framework~\cite{CV-09,CV-10}, the scaling law of the singular part of
the free-energy density around the center of the trap can be written
as
\begin{equation}
F_{\rm sing}({\bf x},T,h) = l^{-3 \theta } 
{\cal F}(rl^{-\theta},\tau l^{\theta y_\tau},hl^{\theta y_h}).
\label{freee}
\end{equation}
where $y_\tau$ and $y_h$ are the RG dimensions reported at the end of
Sec.~\ref{stbh}, and $\theta$ is the trap exponent.  TSS implies that
at the critical point ($\tau=0$) the correlation length $\xi$ of the
critical modes is finite, but increases as $\xi \sim l^{\theta}$ with
increasing the trap size $l$.  The trap exponent can be inferred by a
renormalization-group (RG) analysis of the perturbation induced by the
external trapping potential coupled to the particle density.  We
obtain~\cite{CV-09}
\begin{equation}
\theta = {2\nu\over 1 + 2 \nu}=0.57327(4).
\label{theta}
\end{equation}
The derivation is outlined in App.~\ref{thetaest}.

The TSS equations for the observables and correlation functions
provide an effective description of the critical behavior around the
center of the trap, and, in particular, of the interplay between the
temperature and the confining potential.  At the superfluid transition
and around the center of the trap, the one-particle correlation
function $G_b$ behaves as
\begin{equation}
G_b({\bf x},{\bf y})\approx 
l^{-(1+\eta)\theta} {\cal G}_b({\bf x} l^{-\theta},
{\bf y} l^{-\theta},\tau l^{\theta/\nu}),
\label{twopf}
\end{equation}
and 
the particle-density correlation $G_n$ as
\begin{equation}
G_n({\bf x},{\bf y})\approx 
l^{-2 y_n \theta} {\cal G}_n({\bf x} l^{-\theta},
{\bf y} l^{-\theta},
\tau l^{\theta/\nu}), 
\label{twopfn}
\end{equation}
where $y_n$ is the RG dimension of the density operator
\begin{eqnarray}
y_n=3-1/\nu=1.5112(2). \label{yrho}
\end{eqnarray} 
Analogous scaling relations can be inferred for other correlations.

In our TSS analyses we consider the trap susceptibility $\chi_t$
defined as
\begin{equation}
\chi_t = \sum_{\bf x} G_b({\bf 0},{\bf x}) \label{defchitss}
\end{equation}
(we distinguish it from a generic susceptibility because $\chi_t$ is
the space integral of the correlations with the center of the trap),
and the trap correlation length $\xi_t$ defined from the second moment
of $G_b(0,{\bf x})$, i.e.
\begin{equation}
\xi_t^2 = {1\over 6 \chi_t} \sum_{\bf x} |{\bf x}|^2 G_b(0,{\bf x}) .
\label{defxitss}
\end{equation}
According to TSS, they are expected to behave as
\begin{eqnarray}
&&\chi_t \approx  l^{(2-\eta)\theta} {\cal X}(\tau l^{\theta/\nu})
,\label{chitss}\\
&&\xi_t \approx  l^{\theta} {\cal R}(\tau l^{\theta/\nu}).\label{xitss}
\end{eqnarray}
Note however that any length scale $\xi$ extracted from the critical
modes is expected to show the same TSS as $\xi_t$.

The above TSS equations provide the asymptotic dependence on the trap
size $l$. Scaling corrections are generally expected to be
$O(l^{-\omega\theta})$ where $\omega=0.785(20)$ is the
scaling-correction exponent of the 3D XY universality
class~\cite{CHPV-06,GZ-98,PV-02}, thus
\begin{equation}
\omega \theta = 0.45(1).
\label{omegat}
\end{equation}

We remark that the lattice structure of the BH model does not play any
particular role in the derivation of the TSS formulas, i.e. the
microscopic details of the model are irrelevant in the TSS
limit. Therefore, the above TSS equations apply to a wide class of
models, i.e. to general 3D interacting bosonic systems at the
transition driven by the BEC, thus also including the atomic system
experimentally investigated in Ref.~\onlinecite{DRBOKS-07}.

We also mention that an analogous TSS behavior applies to the 
$T=0$ quantum superfluid-to-Mott phase transition of the 2D BH model
(\ref{bhm}) at fixed integer density~\cite{CV-10}, which belongs to
the same 3D XY universality class~\cite{FWGF-89}.

\section{Finite-size scaling analysis}
\label{cpfss}

Before studying the effect of an external space-dependent trapping
potential, we present a finite-size scaling (FSS) analysis of quantum
Monte Carlo (QMC) simulations of the homogenous BH model (\ref{bhm}),
with periodic boundary conditions.  In particular, we consider the
$U\to\infty$ hard-core limit of the BH model at fixed chemical
potential $\mu=-2$, and vary the temperature along the dashed line
sketched in Fig.~\ref{phdia}.  The QMC simulations are performed using
the stochastic series expansion algorithm with the directed
operator-loop technique~\cite{Sandvik-99,SS-02}.

\subsection{Observables and their FSS}
\label{ssec31}

We compute the particle density $\rho$, the one-particle correlation
function $G_b({\bf x},{\bf y})$,~\cite{footnotegb}
cf. Eq.~(\ref{gbdef}), and the particle-density correlation $G_n({\bf
x},{\bf y})$, cf. Eq.~(\ref{gndef}).  Due to translation invariance,
they only depend on the difference of the arguments, i.e.  $G_\#({\bf
x},{\bf y}) \equiv G_\#({\bf x}-{\bf y})$.

We consider the susceptibility
\begin{equation}
\chi =  \sum_{{\bf x}} G_b({\bf x}),
\label{chisusc}
\end{equation}
which is the zero-momentum component of the Fourier transform
of $G_b$,
\begin{equation}
\widetilde{G}_b({\bf k}) =  \sum_{{\bf x}}
 e^{i{\bf k}\cdot{\bf x}} G_b({\bf x}).
 \label{wgbdef}
\end{equation}
Note that, due to translation invariance, $\widetilde{G}_b({\bf k})$
coincides with the so-called momentum distribution defined by the
double sum (\ref{nkdef}),
apart from a volume factor.  When the system has periodic boundary
conditions, the second-moment correlation length $\xi$ is conveniently
defined by
\begin{equation}
\xi^2 \equiv  {1\over 4 \sin^2 (p_{\rm min}/2)} 
{\widetilde{G}_b({\bf 0}) - \widetilde{G}_b({\bf p})\over 
\widetilde{G}_b({\bf p})},
\label{xidefpb}
\end{equation}
where ${\bf p} = (p_{\rm min},0,0)$, $p_{\rm min} \equiv 2 \pi/L$.  We
also consider the so-called helicity modulus~\cite{CHPV-06,footnoteY}
$\Upsilon$, which is related to the spin stiffness in spin
models~\cite{Sandvik-97}, and to the superfluid density in particle
systems~\cite{FBJ-73,PC-87}.  In QMC simulations using the stochastic
series expansion algorithm, $\Upsilon$ is obtained from the linear
winding number $w_i$ along the $i^{\rm th}$ direction,
\begin{equation}
\Upsilon =  {\langle w_i^2 \rangle\over L},
\qquad w_i = \frac{N_i^+ - N_i^-}{L},
\label{ulw}
\end{equation}
where $N_i^+$ and $N_i^-$ are the number of non-diagonal operators
which move the particles respectively in the positive and negative
$i^{\rm th}$ direction.

FSS predicts the following asymptotic scaling laws of the one-particle
and particle density correlation functions for $r\equiv |{\bf x}|>0$:
\begin{eqnarray}
&&G_b({\bf x}) \approx L^{-1-\eta} {\cal G}_b(r/L,\tau L^{1/\nu}),
\label{gbfss}\\
&&G_n({\bf x}) \approx L^{-2y_n} {\cal G}_n(r/L, \tau L^{1/\nu}),
\label{gnfss}
\end{eqnarray}
where $\tau\equiv T/T_c-1$.
Thus, the susceptibility $\chi$ behaves as
\begin{equation}
\chi = L^{2-\eta}  \left[
g(\tau L^{1/\nu}) + L^{-\omega} g_\omega(\tau L^{1/\nu})+ ...\right],
\label{chil}
\end{equation}
where we have also included the leading $O(L^{-\omega})$ scaling
corrections, and $\omega=0.785(20)$~\cite{CHPV-06,PV-02,GZ-98} is the
critical exponent controlling the leading scaling corrections in the
3D XY universality class.  The dots indicate further scaling
corrections suppressed by higher powers of $1/L$.  The scaling
functions $g$ and $g_\omega$ are universal apart from a multiplicative
constant (since $\chi$ is not RG invariant, $g(0)$ is not universal)
and a rescaling of the argument.

We consider the dimensionless RG invariant quantities
\begin{equation}
R_\xi \equiv \xi/L,\qquad R_\Upsilon \equiv \Upsilon L.
\label{rrr}
\end{equation}
According to the FSS theory~\cite{FBJ-73,Cardy-88,PV-02}, they behave
as (see, e.g., Ref.~\onlinecite{CHPV-06})
\begin{equation}
R = f(\tau L^{1/\nu}) + L^{-\omega} f_\omega(\tau L^{1/\nu})+ ...,
\label{rf}
\end{equation}
around $T_c$ and in the large $L$ limit. $f$ and $f_\omega$ are
scaling functions. In particular the leading one $f$ is universal
(although it depends on the shape of the volume and the choice of the
boundary conditions), i.e. it is independent of the particular model
within the universality class, apart from a trivial rescaling of the
argument.  Thus, $R^*\equiv f(0)$ is universal.  For cubic-shaped
lattices with periodic boundary conditions, the universal
infinite-volume limit of $R_\xi$ and $R_\Upsilon$ at $T=T_c$ are known
with great accuracy:~\cite{CHPV-06} $R_\xi^*=0.5924(4)$ and
$R_\Upsilon^*=0.516(1)$.

\subsection{FSS of QMC data}
\label{ssec32}  

\begin{figure}[tbp]
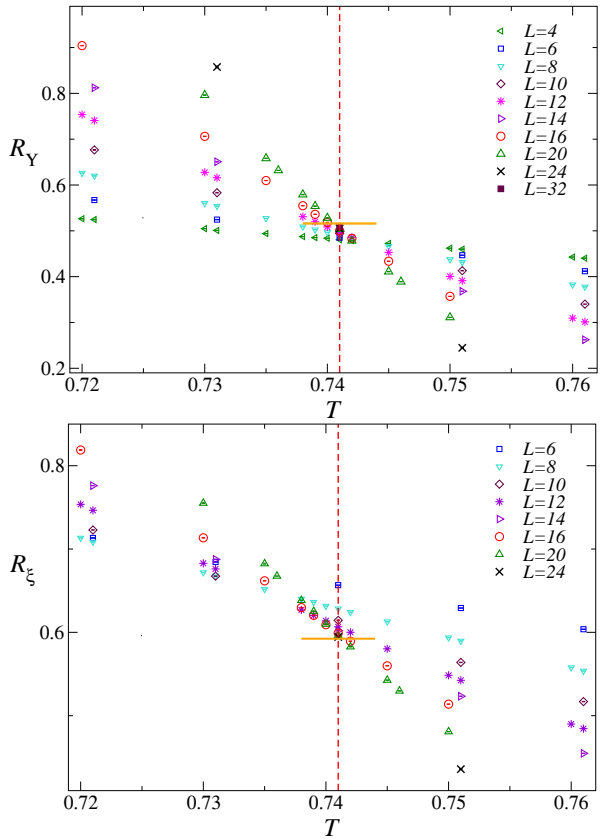

\includegraphics*[scale=\graphicscale]{fig2a.eps}
\includegraphics*[scale=\graphicscale]{fig2b.eps}
\caption{(Color online) QMC data of $R_\xi\equiv \xi/L$ (bottom) and
$R_\Upsilon \equiv \Upsilon L$ (top) for the 3D homogenous BH model
(\ref{bhm}) with periodic boundary conditions.  The vertical dotted
line shows our final estimate of $T_c$, i.e. $T_c=0.7410(1)$.  The
horizontal segments around the crossing point indicate the universal
asymptotic values~\cite{CHPV-06} $R_\xi^*=0.5924(4)$ and
$R_\Upsilon^*=0.516(1)$ at $T_c$.}
 \label{fss}
\end{figure}

\begin{figure}[tbp]
\includegraphics*[scale=\graphicscale]{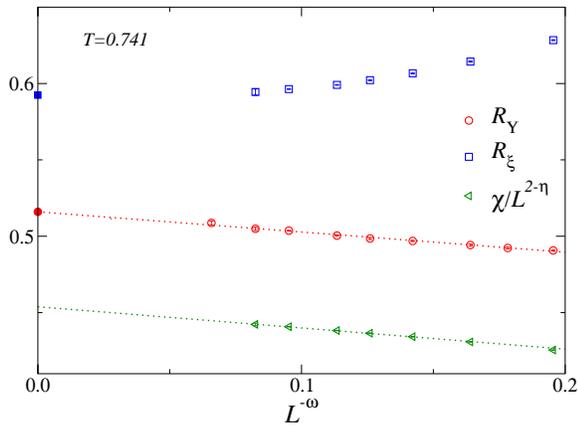}
\caption{(Color online) Data of $R_\Upsilon$, $R_\xi$ and
  $\chi/L^{2-\eta}$ at $T_c$ versus $L^{-\omega}$ with $\omega=0.785$.
  In the case of $R_\Upsilon$ and $R_\xi$ we also show (by full
  symbols) their universal $L\to\infty$ limit: $R_\Upsilon^*=0.516(1)$
  and $R_\xi^*=0.5924(4)$.  The dotted lines show linear fits of the
  data of $R_\Upsilon$ and $\chi/L^{2-\eta}$. In the case of $R_\xi$,
  higher-order scaling corrections appear also significant.  }
 \label{fsstc}
\end{figure}  

\begin{figure}[tbp]
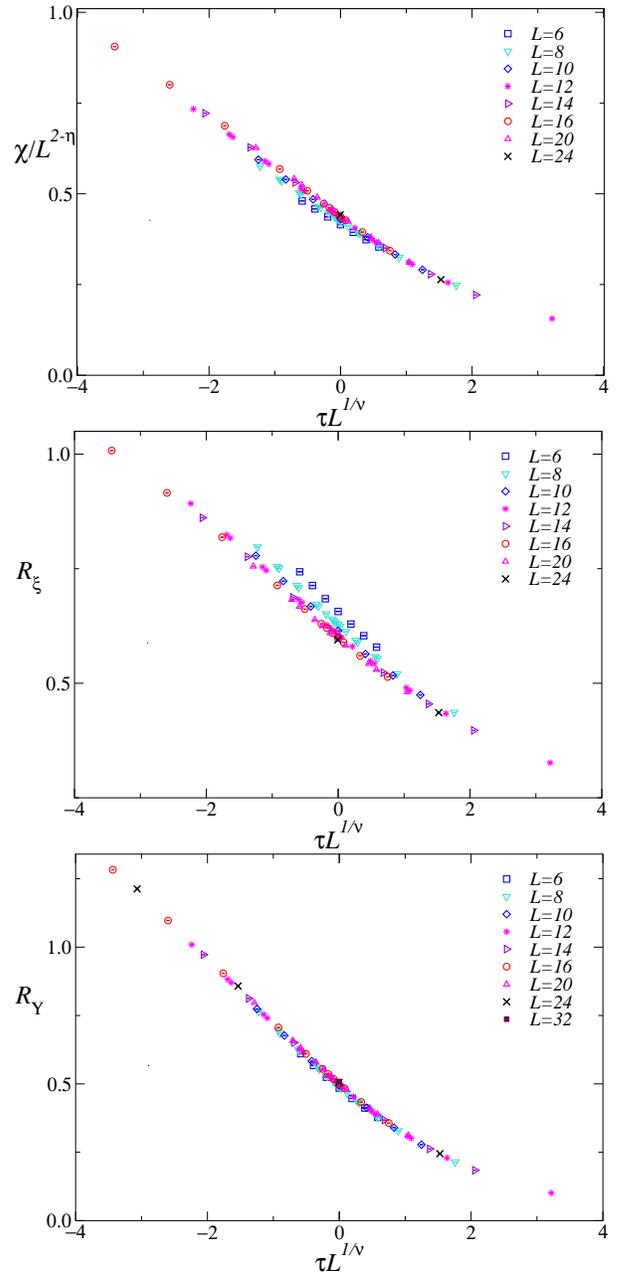

\includegraphics*[scale=\graphicscale]{fig4a.eps}
\includegraphics*[scale=\graphicscale]{fig4b.eps}
\includegraphics*[scale=\graphicscale]{fig4c.eps}
\caption{(Color online) $R_\Upsilon$ (bottom), $R_\xi$ (middle), and
$\chi/L^{2-\eta}$ (top) versus $\tau L^{1/\nu}$ with $\tau\equiv
T/T_c-1$ and $T_c=0.7410$, for homogenous 3D BH model with periodic
boundary conditions }
 \label{fssr}
\end{figure}

We study the FSS of the observables defined above in cubic $L^3$
lattices, up to $L=32$ (up to $L=24$ for observables related to the
one-particle correlation function $G_b$), with periodic boundary
conditions.

In Fig.~\ref{fss} we show the QMC results for $R_\xi$ and
$R_\Upsilon$.  Their sets of data for different lattice sizes show a
clear evidence of a crossing point, whose location is expected to
converge to $T_c$ in the large $L$ limit, according to Eq.~(\ref{rf}).
Moreover, the values of $R_\xi$ and $R_\Upsilon$ at the crossing point
are consistent with the asymptotic universal values $R_\xi^*$ and
$R_\Upsilon^*$ reported above.  The small deviations appear to
decrease with increasing the lattice size; they are explained by the
presence of $O(L^{-\omega})$ corrections, see Eq.~(\ref{rf}) and also
below. An analogous crossing point is shown by the data of the ratio
$\chi/L^{2-\eta}$.

In order to derive an estimate of $T_c$, we fit the data to the Ansatz
\begin{equation}
R = R^* + \sum_{i=1}^n a_i \tau^i L^{i/\nu} + L^{-\omega} \sum_{j=0}^m
b_j \tau^j L^{j/\nu} ,\label{anfit}
\end{equation}
obtained by expanding Eq.~(\ref{rf}) around $\tau=0$.  The best
estimate of $T_c$ is obtained from the data of $R_\Upsilon$.
Sufficiently close to $T_c$, for data with
$|R_\Upsilon/R_\Upsilon^*-1|\lesssim 0.1$ say, the first terms of the
sums [i.e. setting $n=1$ and $m=0$ in Eq.~(\ref{anfit})] provide
already good fits keeping the known universal quantities $\nu$,
$\omega$ and $R_\Upsilon^*$ fixed (in this respect the
$O(L^{-\omega})$ scaling correction term is necessary to achieve fits
with acceptable $\chi^2/{\rm d.o.f}$). For example the fit of the data
for $L\ge 10$ gives $T_c=0.74103(1)$ with $\chi^2/{\rm d.o.f}\approx
1.4$.  We consider
\begin{equation}
T_c=0.7410(1)   \qquad (\mu=-2),
 \label{tcfss}
 \end{equation}
as our final estimate of $T_c$, where the error includes the
statistical errors of the fits, and takes into account the dependence
of the results on the choice of the Ansatz and the interval of values
of $T$ around the transition allowed in the fit. Further subleading
scaling corrections are controlled by increasing the minimum value
$L_{\rm min}$ of $L$ of the data allowed in the fits.  The analysis of
the data of $R_\xi$ gives consistent results, but less precise
because they are apparently affected by larger scaling corrections.

Fig.~\ref{fsstc} shows data of $R_\xi$, $R_\Upsilon$ and
$\chi/L^{2-\eta}$ at $T_c=0.741$, plotted versus $L^{-\omega}$ which
is the expected order of the leading scaling corrections.  As expected
$R_\xi$ and $R_\Upsilon$ converge to their universal values
$R_\Upsilon^*$ and $R_\xi^*$. The approach of $R_\Upsilon$ and
$\chi/L^{2-\eta}$ is approximately linear with respect to
$L^{-\omega}$, while in the case of $R_\xi$ also higher-order scaling
corrections appear significant for the available lattice sizes.

\begin{figure}[tbp]
\includegraphics*[scale=\graphicscale]{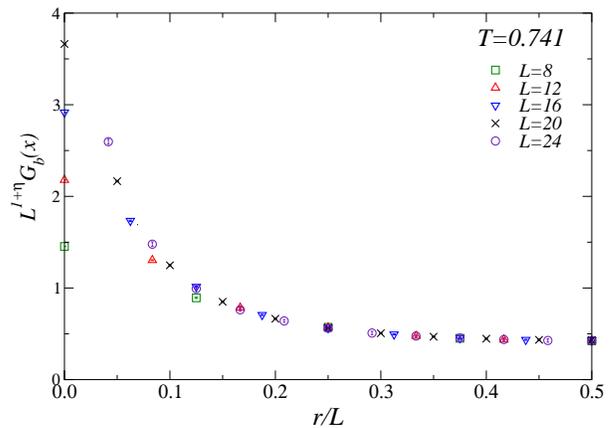}
\caption{(Color online) $L^{1+\eta}G_b({\bf x})$ vs. $r/L$ (where
    $r\equiv|{\bf x}|$) at $T=T_c$ for homogenous BH systems with
    periodic boundary conditions.  The data show the expected scaling
    behavior (\ref{gbfss}).  }
\label{fssgb}
\end{figure}

Fig.~\ref{fssr} reports the QMC data of $R_\Upsilon$, $R_\xi$ and
$\chi/L^{2-\eta}$ versus $\tau L^{1/\nu}$ with $\tau\equiv T/T_c-1$.
They show the asymptotic collapse of the data along a universal curve,
apart from small scaling corrections which get suppressed with
increasing $L$.  Fig.~\ref{fssgb} shows the data of the one-particle
correlation function $G_b$ at $T_c$, which are consistent with the
expected asymptotic scaling behavior reported in Eq.~(\ref{gbfss}).

In conclusion, the above FSS analysis of the QMC data of the 3D
hard-core BH model at $\mu=-2$ definitely confirms that its superfluid
transition belongs to the 3D XY universality class, and provides an
accurate determination of the (nonuniversal) critical temperature,
cf. Eq.~(\ref{tcfss}).

\subsection{Finite-size dependence of the particle density and 
its correlators}
\label{ssec33}  

The behaviors of the particle density, the compressibility, and the
particle-density correlators around the transition are particularly
interesting because they can be directly investigated
experimentally~\cite{FWMGB-06,Campbell-etal-06,
GZHC-09,BGPFG-09,YDCGD-11}, for example by in {\em situ} density image
techniques. Therefore we report a detailed analysis of their data.

\begin{figure}[tbp]
\includegraphics*[scale=\graphicscale]{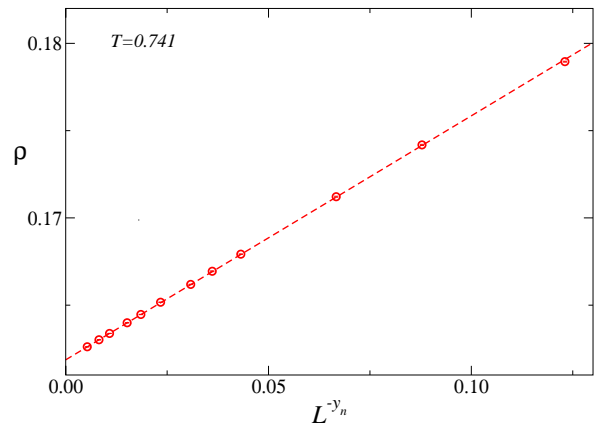}
\caption{(Color online) The particle density at $T_c$ of the 3D
homogenous BH model.  The dashed line shows a linear fit of the data
to $\rho_0 + c \, L^{-y_n}$, with $\rho_0=0.16187(1)$ and
$c=0.140(1)$.  }
 \label{fssdetc}
\end{figure}

The periodic boundary conditions preserve translation invariance,
which implies that the particle density $\rho({\bf x})$ is independent
of ${\bf x}$, thus
\begin{equation}
\rho({\bf x}) \equiv 
\rho = {1\over L^3} \langle \hat{N}\rangle, \qquad 
\hat{N}=\sum_{\bf x} n_{\bf x}.
\label{rhocpb}
\end{equation}
Its behavior around the transition point is analogous to that of the
energy density in spin systems, see e.g. Ref.~\onlinecite{CHPV-06},
thus
\begin{equation}
\rho \approx   f_a(\tau) + L^{-y_n} f_s(\tau L^{1/\nu})  
\label{rhofss2}
\end{equation}
at fixed chemical potential, where $f_a$ is a nonuniversal
analytic function of $\tau$ (and $\mu$), $y_n$ is the RG dimension of
the particle density operator $n_x$, cf. Eq.~(\ref{yrho}), and
$f_s$ is a universal function apart from a factor and a rescaling
of its argument. This behavior is clearly shown by the data at $T_c$,
which are well approximated by the asymptotic formula
\begin{equation}
\rho \approx \rho_0 + c \, L^{-y_n},
\label{rhoc}
\end{equation}
as shown by Fig.~\ref{fssdetc}.
A linear fit to (\ref{rhoc}) gives $\rho_0 = 0.16187(1)$.  

\begin{figure}[tbp]
\includegraphics*[scale=\graphicscale]{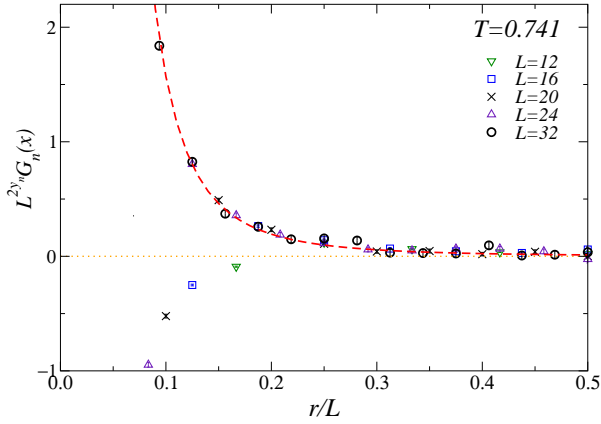}
\caption{(Color online) $L^{2y_n}G_n(x)$ vs. $r/L$ at $T=T_c$ for
  homogenous systems with periodic boundary conditions.  The dashed
  line sketches the expected asymptotic behavior $G_n({\bf x})\sim
  r^{-2y_n}$ at small $r\equiv|{\bf x}|$.  }
\label{fssgn}
\end{figure}

\begin{figure}[tbp]
\includegraphics*[scale=\graphicscale]{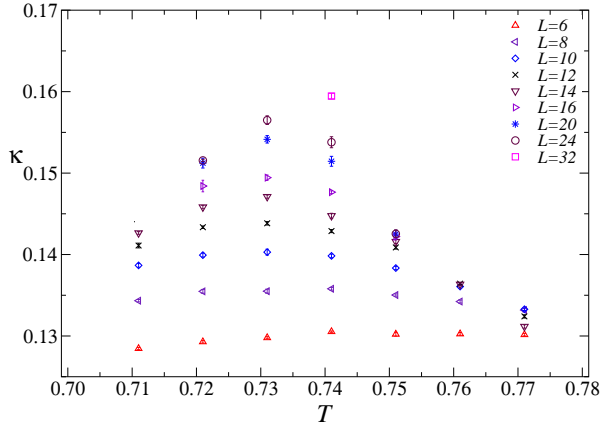}
\caption{(Color online) 
QMC data of the compressibility $\kappa$, cf.  Eq.~(\ref{vdefss}).
}
 \label{fssfluvst}
\end{figure}

\begin{figure}[tbp]
\includegraphics*[scale=\graphicscale]{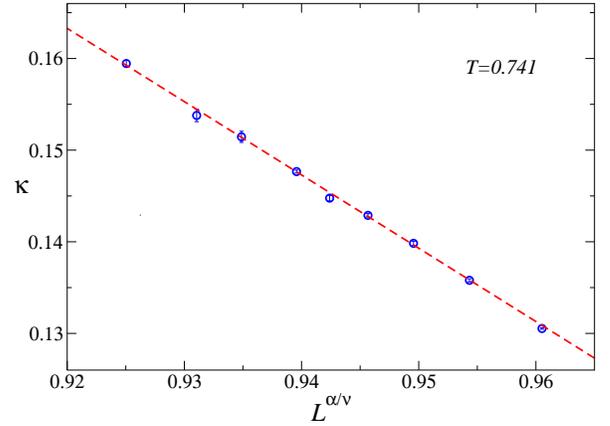}
\caption{(Color online) QMC data of the compressibility $\kappa$ at
  $T_c$.  The dashed line shows a linear fit to the predicted
  asymptotic behavior $ a + b L^{\alpha/\nu}$; in particular, the data
  for $L\ge 8$ give $a=0.90(1)$ and $b\approx -0.80(1)$ with
  $\chi^2/{\rm d.o.f.}\approx 1.1$.  }
 \label{fssflu}
\end{figure}

Fig.~\ref{fssgn} reports data of the particle-density correlation
function at $T_c$.  They show the expected scaling behavior,
obtainable from Eq.~(\ref{gnfss}) setting $\tau=0$. Note that it
develops for positive values of $G_n$, while the negative data at
small distance are pushed toward the origin, not contributing 
to the scaling behavior. 
The space integral of $G_n$ gives the compressibility
\begin{equation}
\kappa \equiv  {\partial \rho\over \partial \mu} =
\sum_{\bf x} G_n({\bf x}) = {1\over L^3} 
\left( \langle \hat{N}^2 \rangle - \langle \hat{N}\rangle^2\right).
\label{vdefss}
\end{equation}
Its scaling behavior is complicated by the sum around ${\bf x}=0$,
which gives rise to a nonuniversal analytic contribution, analogously
to the specific heat in $^4$He, see e.g. Ref.~\onlinecite{PV-02}.
Indeed, we expect
\begin{equation}
\kappa \approx g_a(\tau) + L^{\alpha/\nu} g_s(\tau L^{1/\nu}), 
\label{vdefss2}
\end{equation}
where $\alpha$ is the specific heat exponent
$\alpha=-0.0151(3)$,~\cite{CHPV-06} $g_a$ is a nonuniversal analytic
function of $\tau$, and $g_s$ is a universal function apart from a
factor and a rescaling of the argument.  Notice that, since
$\alpha<0$, the nonuniversal analytic term provides the leading
behavior for $L\to\infty$.  Fig.~\ref{fssfluvst} shows the data of the
compressibility. They hint at the typical $\lambda$ shape expected in
the infinite-volume limit, which also characterizes the specific heat
at the superfluid transition of $^4$He, see
e.g. Refs.~\onlinecite{Lipa-etal-96,GKMD-08}.  At $T_c$ they show the
asymptotic scaling behavior
\begin{equation}
\kappa  = a + b \,L^{\alpha/\nu},
\label{vtc}
\end{equation}
see Fig.~\ref{fssflu}.  Linear fits of the available data at $T_c$, up
to $L=32$, give $a\approx 0.90$ and $b\approx -0.80$.

\section{Critical parameters from TSS}
\label{critss}

We now consider the 3D BH model in the presence of a trapping
potential, as described by the Hamiltonian (\ref{bhmt}).  We present
results of QMC simulations of the 3D hard-core BH model at $\mu=-2$
for several values of the trap size $l$, up to $l=14$.  The trap is
centered in the middle of a cubic $L^3$ lattice, with odd $L$ and open
boundary conditions. More details on the practical implementation of
QMC simulations of trapped systems can be found in
Refs.~\onlinecite{CTV-12,CT-12}.

The lattice size $L$ is taken sufficiently large to effectively
reproduce the infinite-volume limit, i.e., so that the residual
finite-size effects can be considered negligible compared with the
statistical errors, at least for the critical correlations around the
trap.  This is checked by comparing results at fixed trap size $l$
with increasing the lattice size $L$.  In particular, after some
checks, simulations up to $l=10$ were generally performed taking
$L=4l+1$ (we mention that for $l=10$ and $L=4l+1$, the particle
density is $\rho\approx 0.157$ at the center of the trap, and
$\rho\lesssim 0.001$ at the boundaries), while for those at $l=14$ we
considered $L/l^\theta\gtrsim 7$.  We return to this point below.

\subsection{TSS analyses of QMC data}
\label{tctsssec}

We now show that a TSS analysis of the data for trapped systems allows
us to determine the critical parameters, analogously to the FSS
analysis presented in the previous section.

\begin{figure}[tbp]
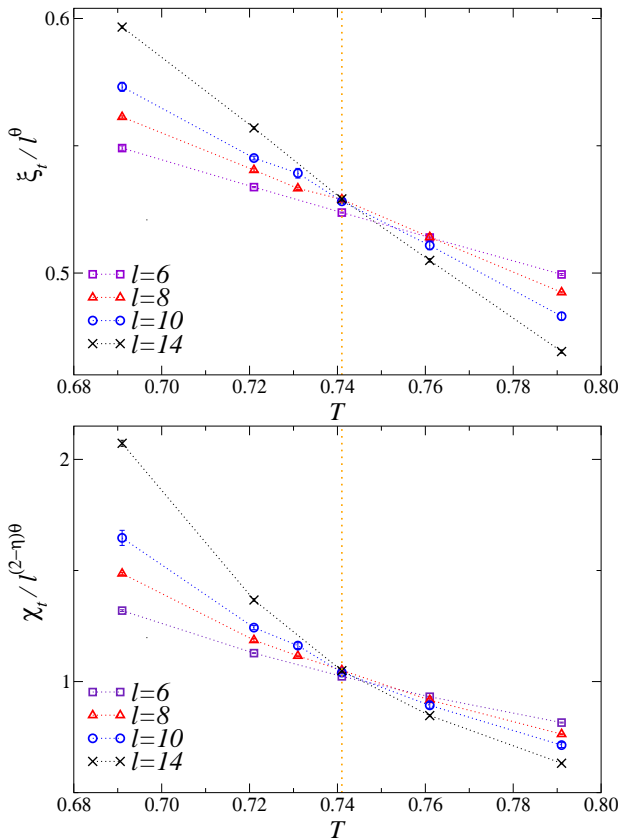

\includegraphics*[scale=\graphicscale]{fig10a.eps}
\includegraphics*[scale=\graphicscale]{fig10b.eps}
\caption{(Color online) QMC data of $\xi_t/l^\theta$ (top) and
  $\chi_t/l^{(2-\eta)\theta}$ (bottom) in the presence of the trap.
  The vertical dotted line indicates the critical temperature
   $T_c=0.7410$ obtained from
  the FSS analysis of Sec.~\ref{ssec32}.  
  The dotted lines connecting the data are drawn to guide the eyes.}
\label{tss}
\end{figure}

\begin{figure}[tbp]
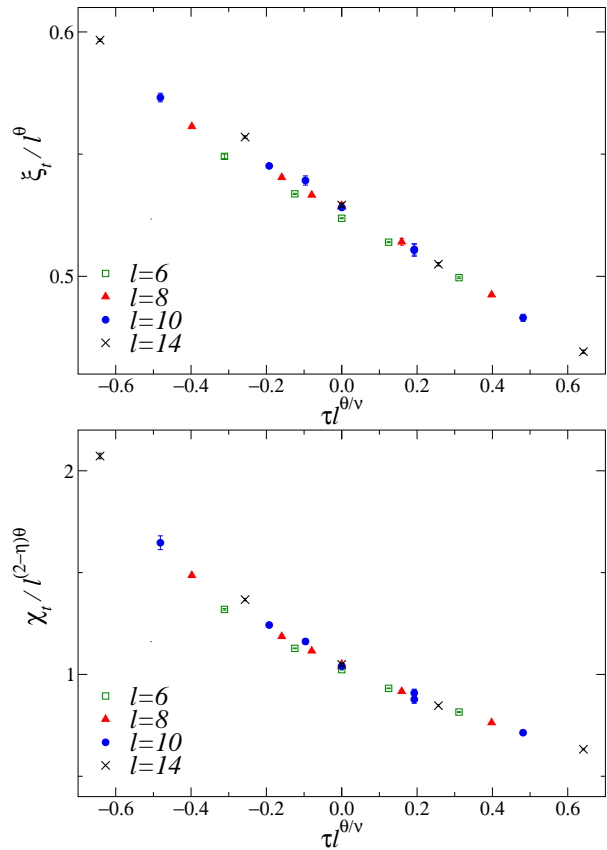

\includegraphics*[scale=\graphicscale]{fig11a.eps}
\includegraphics*[scale=\graphicscale]{fig11b.eps}
\caption{(Color online) QMC data of $\xi_t/l^\theta$ (top) and
  $\chi_t/l^{(2-\eta)\theta}$ (bottom) in the presence of the trap vs
  $\tau l^{\theta/\nu}$ with $\tau\equiv T/T_c-1$ and $T_c=0.741$.  }
\label{tssr}
\end{figure}

In order to determine $T_c$ from TSS, cf. Eqs.~(\ref{chitss}) and
(\ref{xitss}), we may exploit the fact that at $T=T_c$
(i.e. $\tau=0$), the ratios $\chi_t/l^{\theta(2-\eta)}$ and
$\xi_t/l^\theta$ become independent of the trap size $l$ in the
large-$l$ limit.  Therefore, we expect that sets of data for different
trap sizes cross each other at one value of the temperature (apart
from scaling corrections), providing an estimate of $T_c$, analogously
to the FSS analysis of the previous section.  This is indeed observed
in Fig.~\ref{tss}, which shows the available data of
$\chi_t/l^{\theta(2-\eta)}$ and $\xi_t/l^\theta$ versus $T$.  The
apparent crossing point of the TSS data indicates $T_c\approx 0.74$.
A more accurate estimate is achieved by fitting the data to
the simple Ansatz
\begin{equation}
a + b (T-T_c) l^{\theta/\nu},
\label{tctssan}
\end{equation}
considering data sufficiently close to the crossing point to avoid
higher powers of $(T-T_c) l^{\theta/\nu}$ (the optimal region turns out
to be $0.72\lesssim T \lesssim 0.76$).  We obtain $T_c=0.741(2)$ and
$T_c=0.742(2)$ respectively from the data of $\xi_t/l^\theta$ and
$\chi_t/l^{(2-\eta)\theta}$ (in both cases from data for $l\ge 8$).
The expected $O(l^{-\omega\theta})$ scaling corrections,
cf. Eq.~(\ref{omegat}), turn out to be negligible, at least for $l\ge 8$
and close to the crossing point.

\begin{figure}[tbp]
\includegraphics*[scale=\graphicscale]{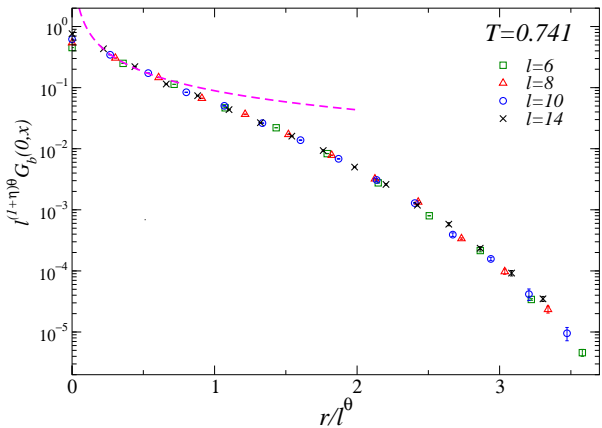}
\caption{(Color online) The one-particle correlation function
  $G_b(0,{\bf x})$ at $T_c$.  The dashed line shows the expected
  small-distance behavior $G_b(0,{\bf x})\sim
  r^{-(1+\eta)}$.}
\label{tssbb}
\end{figure}

In Fig.~\ref{tssr} we plot the data of of $\chi_t/l^{\theta(2-\eta)}$
and $\xi_t/l^\theta$ versus $\tau l^{\theta/\nu}$,
with $\tau\equiv T/T_c-1$ and $T_c=0.741$.
They are consistent
with the scaling behavior predicted by Eqs.~(\ref{chitss}) and
(\ref{xitss}), approaching a universal curve in the large-$l$ limit.
Deviations are expected to be $O(l^{-\omega\theta})$,
i.e.  they get suppressed as $l^{-0.45}$.
The TSS of the one-particle correlation function $G_b(0,{\bf x})$ at
$T_c$, i.e.
\begin{equation}
G_b(0,{\bf x}) =  l^{-(1+\eta)\theta} g_b(X), \quad X\equiv r/l^\theta,
\label{twopf2}
\end{equation}
is nicely reproduced by the data shown in Fig.~\ref{tssbb}.  At small
distance the two-point function $G_b$ is expected to show the
power-law behavior of the homogenous system, i.e.  $G_b(0,{\bf x})
\sim {1/r^{1+\eta}}$.

In conclusion, although the available data come from moderately large
trap sizes, up to $l=14$,
their analysis shows a clear evidence of the expected TSS,
which allows us to accurately estimate the critical temperature $T_c$
with a precision of a few per mille, in good agreement with the
estimate of $T_c$ by the standard FSS analysis of Sec.~\ref{ssec32},
i.e. $T_c=0.7410(1)$.
Let us also mention that an independent estimate of $\theta$ can be
obtained by a nonlinear fit of the data of $\xi_t$ to
\begin{equation}
\xi_t = l^\theta \left[ a + b(T-T_c) l^{\theta/\nu}\right],
\label{xitansatztheta}
\end{equation}
and analogously for $\chi_t$.  Fixing $T_c=0.7410$, these fits
give  $\theta
\approx 0.58$ from the data around $T_c$ (and $l\ge 8$), in agreement
with the TSS result (\ref{theta}). However, the available
data of $\xi_t$ and $\chi_t$
do not allow us to obtain sufficiently robust results
from more general fits leaving free also $T_c$.  More data around the
critical point and, more importantly, for larger trap sizes are
necessary for this purpose.

\subsection{Finite-size effects with the trap}
\label{ivlim}

\begin{figure}[tbp]
\includegraphics*[scale=\graphicscale]{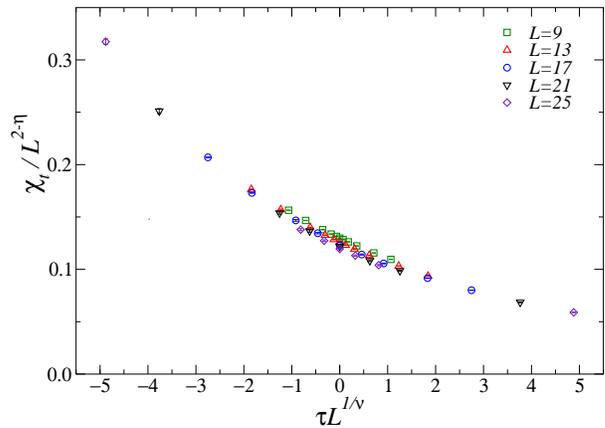}
\caption{(Color online) Finite-size scaling of the trap dependence of
  the QMC data of $\chi/L^{2-\eta}$ in the presence of the trap, from
  QMC simulations keeping $L/l^\theta=2$ fixed and using open boundary
  conditions.  }
\label{ftssr}
\end{figure}

The finite-size effects in the presence of the trap, i.e. when
considering the trap within a finite box of size $L$ (with open
boundary conditions), can be taken into account by adding a further
dependence on $L l^{-\theta}$ in the TSS Ansatz~\cite{QSS-10}
(\ref{freee}), and in Eqs.~(\ref{twopf}) and (\ref{twopfn}) as well.  For
example the finite-size and trap-size scaling (FTSS) of the trap
susceptibility $\chi_t$ and the correlation length $\xi_t$,
cf. Eqs.~(\ref{defchitss}) and (\ref{defxitss}), can be written as
\begin{eqnarray}
&&\chi_t \approx L^{2-\eta} {\cal X}(\tau l^{\theta/\nu},L/l^\theta),
\label{chiftss}\\
&&\xi_t \approx  L {\cal R}(\tau l^{\theta/\nu},L/l^\theta).
\label{xiftss}
\end{eqnarray}
The above scaling is confirmed by the data of $\chi_t$ shown in
Fig.~\ref{ftssr}, obtained by QMC simulations keeping $L/l^\theta=2$
fixed.

\begin{figure}[tbp]
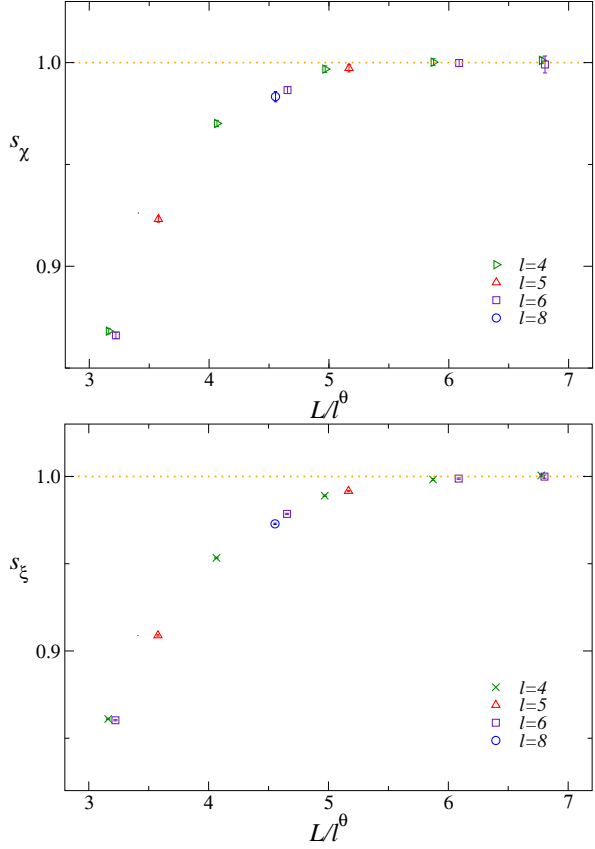

\includegraphics*[scale=\graphicscale]{fig14a.eps}
\includegraphics*[scale=\graphicscale]{fig14b.eps}
\caption{(Color online) Finite-size scaling curves of the trap
  susceptibility $\chi_t$ and correlation length $\xi_t$ at $T_c$,
   cf. Eqs.~(\ref{chiftss2}) and~(\ref{xiftss2}).  }
\label{ftsscurves}
\end{figure}

FTSS also implies that at $T_c$ the ratio of quantities computed in
box of size $L$ becomes only function of $L/l^\theta$ asymptotically,
i.e.,
\begin{eqnarray}
&&s_\chi\equiv {\chi_t(l,L)\over \chi_t(l,L\to\infty)} = f_\chi(L/l^\theta) ,
\label{chiftss2}\\
&&s_\xi\equiv {\xi_t(l,L)\over \xi_t(l,L\to\infty)} = f_\xi(L/l^\theta) .
\label{xiftss2}
\end{eqnarray}
Their data at $T_c$ support this scaling behavior, see
Fig.~\ref{ftsscurves}.  

Moreover they tell us that around the transition the finite-size
effects on $\chi_t$ and $\xi_t$ get smaller than 0.1\% when
$L/l^\theta \gtrsim 7$.  All data reported in Sec.~\ref{tctsssec},
which were supposed to correspond to the infinite size limit, were
obtained by simulations of systems satisfying this condition.

\subsection{Trap-size dependence of the particle density}
\label{pdtss}

Finally, we discuss the trap-size dependence of the particle density,
which has been considered in the literature as a possible probe
of critical behavior, due to the experimental capability of measuring
it quite accurately.

Analogously to the FSS of homogenous systems, the scaling behavior of
the particle density is more involved, because it is dominated by an
analytical (non critical) contribution.  In the presence of the trap,
its behavior is further complicated by the fact that the particle
density depends on the distance from the center of the trap.

\begin{figure}[tbp]
\includegraphics*[scale=\graphicscale]{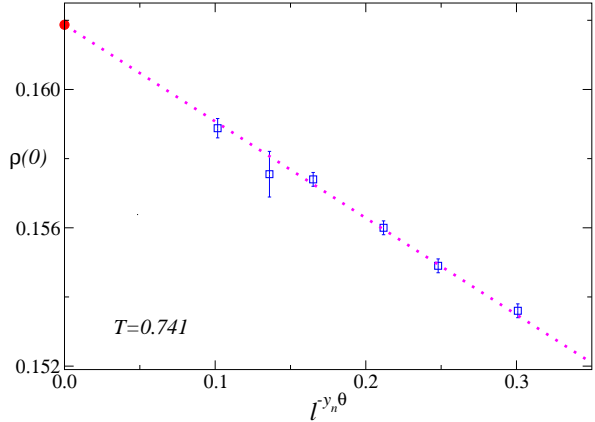}
\caption{(Color online) The particle density at $T_c$ and at the
  center of the trap.  The full circle along the $y$-axis shows the
  value $\rho_0=0.16187$ of the leading asymptotic term obtained for
  homogenous systems, see Fig.~\ref{fssdetc}.  The dotted line shows a
  linear fit to $\rho_0 + b \, l^{-\theta y_n}$, which gives
  $\rho_0=0.1617(3)$ and $b=-0.027(1)$.  }
\label{tssde0}
\end{figure}

To begin with, we consider the particle density at the center of the
trap, ${\bf x}=0$. Its asymptotic trap-size dependence is expected to
be
\begin{equation}
\rho(0) =   g_a(\tau) + l^{-y_n\theta} g_s(\tau l^{\theta/\nu})  + ...,
\label{rhotss0}
\end{equation}
where $y_n\theta=0.8664(3)$, $g_a$ is a nonuniversal analytical
function, and $g_s$ is a scaling function. Moreover, the local-density
approximation (LDA), see e.g. Refs.~\onlinecite{BDZ-08,CTV-12},
suggests that the leading analytical term $g_a(\tau)$ is identical to
that of Eq.~(\ref{rhofss2}) for homogenous systems.  This is supported
by the data at $T_c$ shown in Fig.~\ref{tssde0}, which are consistent
with the asymptotic formula
\begin{equation}
\rho(0) \approx \rho_0 + b \, l^{-y_n\theta},
\label{rhoctss}
\end{equation}
with $\rho_0$ equal to the leading constant term of homogenous systems,
cf.  Eq.~(\ref{rhoc}) with $\rho_0=0.16187(1)$. Indeed, a linear fit
of the data to Eq.~(\ref{rhoctss}) gives $\rho_0=0.1617(3)$ and
$b=-0.027(1)$ with $\chi^2/{\rm d.o.f.}\approx 0.4$.

\begin{figure}[tbp]
\includegraphics*[scale=\graphicscale]{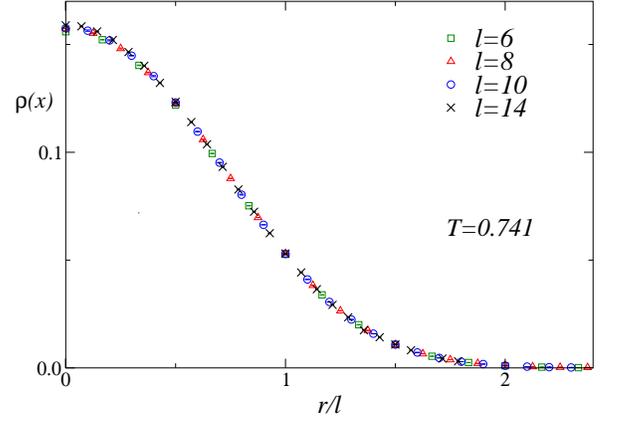}
\caption{(Color online) The space dependence of the particle 
density $\rho(x)$ at $T_c$ in the presence of the trap.  }
\label{tssde}
\end{figure}

Concerning the space dependence of the particle density, and using
rotational invariance, we expect that its large trap-size behavior is
\begin{equation}
\rho({\bf x}) \approx  f_a(r/l,T)  + 
l^{-y_n\theta} f_s(r/l^\theta,\tau l^{\theta/\nu}),
\label{rhoxtss3}
\end{equation}
where $f_a$ is again an analytic function. Its analytic dependence on
the ratio $r/l$ is quite natural, because we expect it to be a smooth
function of
\begin{equation}
\mu_{\rm eff} (r/l)\equiv \mu - V(r) = \mu - (r/l)^2,
\label{mueff}
\end{equation}
as suggested by LDA.  The asymptotic dependence on $r/l$ is shown by
the plot of Fig.~\ref{tssde}.  We expect that the leading analytic
function $f_a(r/l,T)$ is provided by the LDA approximation, i.e. by
the particle density $\rho_h(\mu_{\rm eff},T)$ of the homogenous
system in the infinite-volume limit.  The asymptotic validity of the
LDA of the particle density was also found at the $T=0$ quantum
transitions of 1D and 2D BH models~\cite{CV-10b,CTV-12,CT-12}.

\begin{figure}[tbp]
\includegraphics*[scale=\graphicscale]{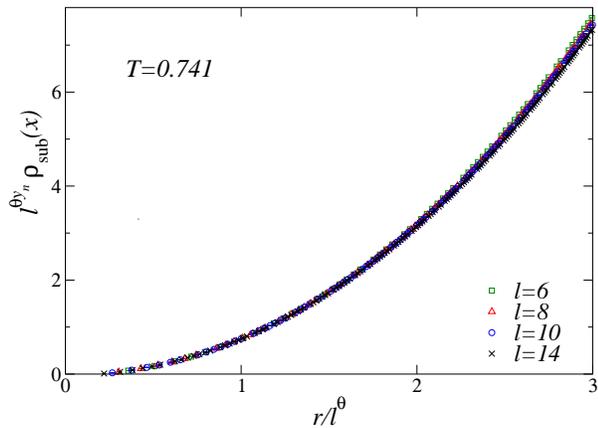}
\caption{(Color online) The subtracted particle density $\rho_{\rm
sub}(x)$ at $T_c$, cf Eq.~(\ref{rhosub}).  The data of
$l^{y_n\theta}\rho_{\rm sub}$ versus $r/l^\theta$ for different trap
sizes collapse toward a unique curve, confirming the scaling behavior
(\ref{rhosub}).  }
\label{tssde3}
\end{figure}

The above results show that the behavior of the particle density
around the center of the trap and across the transition is quite
nontrivial. We finally write it as
\begin{equation}
\rho({\bf x}) = 
\rho_h[\mu_{\rm eff}(r/l),T] + 
l^{-y_n\theta} f_s(r/l^\theta,\tau l^{\theta/\nu}).
\label{rhoxtss2}
\end{equation}
Let us consider the TSS limit at $T=T_c$ of this asymptotic behavior,
i.e. $l\to\infty$ keeping $X\equiv r/l^\theta$ fixed.  Since in this
limit $r/l=X/l^{1-\theta}\to 0$, we can expand the analytical term,
obtaining
\begin{eqnarray}
\rho({\bf x}) &=& \rho_h(\mu) - l^{-2(1-\theta)} \kappa_h(\mu) X^2 
+\, O(l^{-4(1-\theta)}) \nonumber \\
&+&  l^{-y_n\theta} f_s(X) + O(l^{-(y_n+\omega)\theta}),
\label{rhotss3}
\end{eqnarray}
where $\kappa_h(\mu)\equiv {\partial\rho_{\rm
h}(\mu)/\partial\mu}$.  Note that the $O(l^{-2(1-\theta)})$ term
cannot be neglected with respect to the scaling term, because
\begin{equation}
2(1-\theta)<y_n\theta<4(1-\theta).
\end{equation}
Indeed, $2(1-\theta)=0.8535(1)$ and $y_n\theta=0.8664(3)$.  Therefore,
in order to determine the universal scaling term, we must subtract the
terms containing $\rho_h(\mu)$ and $\kappa_h(\mu)$. They may be
evaluated from calculations within the homogenous model at $\mu$ and
$T$ fixed, for example considering their large-$L$ limit using
periodic boundary conditions.  Using the corresponding results at
$T_c$ for the FSS of homogenous systems, see Sec.~\ref{cpfss}, we
estimate $\rho_h(\mu=-2,T_c)\approx 0.16187$ and
$\kappa_h(\mu=-2,T_c)\approx 0.90$.  Then we define the subtracted
particle density
\begin{eqnarray}
\rho_{\rm sub}({\bf x}) &\equiv& \rho({\bf x}) - \rho_h(\mu,T) 
+ l^{-2(1-\theta)} 
\kappa_h(\mu,T) X^2 \nonumber \\
&\approx&
l^{-y_n\theta} f_s(X), \qquad X\equiv r/l^\theta.
\label{rhosub}
\end{eqnarray}
Its scaling behavior is nicely confirmed by the corresponding data
plotted in Fig.~\ref{tssde3}.

\begin{figure}[tbp]
\includegraphics*[scale=\graphicscale]{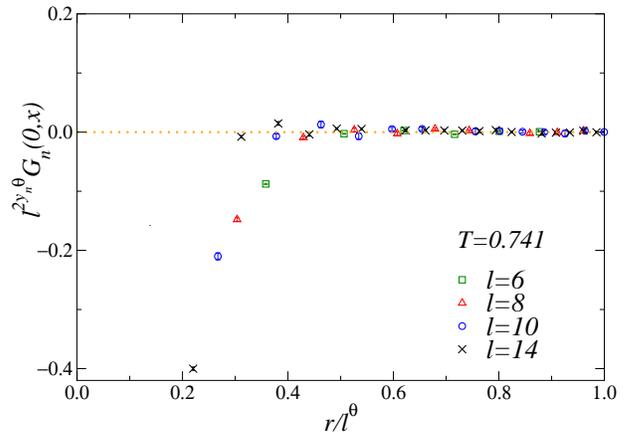}
\caption{(Color online) The density-density correlation $G_n(0,x)$ 
at $T_c$.}
\label{tssgn}
\end{figure}

Finally in Fig.~\ref{tssgn} we show our data for the particle-density
correlation $G_n$, which vanish at relatively small distance,
and  do not apparently show scaling behaviors,
likely because the trap size of the available data is still too small.
Indeed the FSS data of Fig.~\ref{fssgn} begin showing scaling at
relatively large values of the size, essentially because the
correlation function is significantly nonzero only at small distance.

\section{Conclusions}
\label{conclusions}

We investigate the critical behavior of trapped particle systems at
the finite-temperature superfluid transition driven by the formation
of a Bose-Einstein condensate (BEC).  In particular, we consider lattice
particle systems described by the 3D Bose-Hubbard (BH) model
(\ref{bhmt}), which is a realistic model of cold bosonic atoms in
optical lattices~\cite{JBCGZ-98,BDZ-08}.  We present FSS and TSS
analyses of numerical QMC simulations of the homogenous and trapped 3D
BH model in the hard-core $U\to\infty$ limit at fixed $\mu=-2$.

We show that an accurate study of the critical behavior, and accurate
determinations of the critical parameters, can be achieved by matching
the trap-size dependence of some appropriate observables with the
scaling predicted by trap-size scaling (TSS), see
e.g. Eqs.~(\ref{twopf}) and (\ref{xitss}).  The main advantage of this
approach is that it is supposed to exactly converge to the critical
parameters in the large trap-size limit, thus providing a systematic
scheme to improve the results and control the uncertainty, without
using further assumptions and approximations, such as
mean-field and local-density approximation (LDA).  Although the
available QMC data are obtained for moderately large trap sizes, our
TSS analysis allows us to accurately estimate the critical temperature
$T_c$, e.g.  $T_c=0.741(2)$ from the analysis of the trap correlation
length, cf. Eq.~(\ref{defxitss}),
which agrees with the critical value $T_c=0.7410(1)$ obtained
by a standard FSS analysis of QMC simulations of the homogenous 3D
BH model.

Our numerical analysis may provide a guide for experimental
investigations of finite-temperature and quantum transitions in
physical systems that are made inhomogenoues by the presence of an
external space-dependent field, like the experimentally interesting
case of trapped  atomic systems~\cite{CWK-02,BDZ-08,GPS-08}.  
Indeed, it should 
show how the critical parameters may be determined by looking
at the scaling of the critical modes with respect to the trap size,
i.e.  by matching the trap-size dependence of the experimental data
with the expected TSS Ansatz, similarly to experiments probing the
finite-size scaling behavior of homogenous $^4$He systems at the
superfluid transition~\cite{GKMD-08}.

A few comments are in order concerning the optimal observables to
determine the critical parameters in trapped systems.  Of course, they
are those which are closely related to the critical modes around
the center of the trap. In particular, the most convenient quantities
are those whose leading behavior in the large trap-size limit is given
by the universal TSS associated with the critical modes.  In the case
of the superfluid transition or BEC, the optimal quantities are
related to the one-particle correlation function of the bosonic field
around the center ${\bf x}=0$ of the trap, which scales as $G_b(0,{\bf
x})\approx l^{-(1+\eta)\theta} {\cal G}_b(rl^{-\theta},\tau
l^{\theta/\nu})$.  For example, we consider the trap susceptibility
$\chi_t=\sum_{\bf x} G_b(0,{\bf x})$ and the trap correlation length
$\xi_t$ defined from the second moment of $G_b(0,{\bf x})$, which
scale as $\chi_t \approx l^{(2-\eta)\theta} {\cal X}(\tau
l^{\theta/\nu})$ and $\xi_t = l^{\theta} {\cal R}(\tau
l^{\theta/\nu})$, respectively.  Note that any length scale $\xi$
extracted from the critical modes is expected to show the same TSS
behavior as $\xi_t$, and therefore to be effective to determine the
critical parameters.

Since the particle density can be accurately measured in experiments
of trapped cold atoms, for example by {\em in situ} density image
techniques~\cite{GZHC-09,BGPFG-09,HZHTGC-11}, its behavior may be used
as a probe of the critical behavior at finite-temperature and quantum
transitions, see e.g.
Refs.~\onlinecite{ZKKT-09,ZH-10,PPS-10,FCMCW-11,ZHTGC-11,HM-11,CR-12}.
The particle density turns out to show a more involved scaling
behavior, given by Eq.~(\ref{rhoxtss3}).  In this case the nontrivial
TSS arising from the critical modes, which contains the information on
the critical behavior, does not provide the leading contribution to
the particle density. Indeed, it scales as $O(l^{-y_n\theta})$ with
$y_n\theta\approx 0.866$, while the dominant contribution in the large
trap-size limit is given by an analytical background function
$f_a(r/l,T)$ related to noncritical modes, cf. Eq.~(\ref{rhoxtss3}).  
Our numerical analysis,
see Sec.~\ref{pdtss}, shows that such leading analytical background is
well approximated (actually we conjecture that it is exactly given) by
the corresponding LDA, i.e. by the infinite-volume limit of homogenous
systems at chemical potential $\mu_{\rm eff}= \mu-V(r)$. Thus, TSS
provides the leading behavior of the deviations from the LDA around
the transition.  Therefore, a careful subtraction is required in order
to observe the genuine critical term which carries information on the
critical temperature and the critical exponents.  As a consequence, in
order to infer the critical parameters from the particle density, very
accurate experimental data are required to fit them to
Eq.~(\ref{rhoxtss3}).  Alternatively, one may resort to some
approximations for the analytical background, but this may make the
control of the real uncertainty more questionable, significantly
affecting the accuracy of the results.

The particle-density correlations and compressibility are also working
optical lattice observables, see
e.g. Refs.~\onlinecite{FGWMGB-05,SPP-07,HZHTGC-11}.  An analogous TSS
applies to the connected particle-density correlation (\ref{gndef}),
for example $G_n(0,{\bf x})\approx l^{-2\theta y_n } {\cal
G}_n(rl^{-\theta},\tau l^{\theta/\nu})$.  Thus information on the
critical behavior can be achieved by matching the experimental data to
this TSS Ansatz.  However, our numerical analysis shows that the
universal scaling of $G_n$ at the superfluid transition turns out to
appear at relatively small distance, see Fig.~\ref{fssgn} and
\ref{tssgn}, which may make an accurate determination of its scaling
quite hard.

As already mentioned, a promising quantity is the trap susceptibility
$\chi_t\equiv \sum_{\bf x} G_b({\bf 0},{\bf x}) \sim l^{(2-\eta)\theta}$
with $(2-\eta)\theta\approx 1.12$, cf. Eqs.~(\ref{defchitss}) and
(\ref{chitss}). However, we should note that $\chi_t$ is not
proportional to the zero-momentum component of the momentum
distribution, which is given by $n({\bf k})\equiv \sum_{{\bf x},{\bf
y}} e^{i{\bf k}\cdot({\bf x}-{\bf y})} G_b({\bf x},{\bf y})$ even in
the presence of the trap, and which can be experimentally related to
the interference patterns of absorption images after a time-of-flight
period in the large-time ballistic regime, see
e.g. Ref.~\onlinecite{BDZ-08}.  Simple considerations show that 
$n({\bf k})$, and in particular its
zero-momentum component, is largely dominated by the noncritical
regions of the trap, while the contribution of the critical modes are
suppressed, roughly by a total volume factor of the system.  Its
critical scaling in trapped system in not clear, at least in the TSS
framework, thus the global momentum distribution of the system does
not appear promising to accurately determine the critical parameters.
See, however, Refs.~\onlinecite{DZZH-07,Trotzky-etal-10,PPS-10,CR-12}
for a discussion of methods based on the measurement of the momentum
distribution.

We remark that the above considerations also apply to generic trapped
interacting bosonic particle systems at the transition driven by 
BEC, such as the atomic system experimentally
investigated in Ref.~\onlinecite{DRBOKS-07}.

Let us finally mention that a substantial different TSS is expected in
2D trapped bosonic systems.  The finite-temperature superfluid
transition of 2D interacting bosonic systems is described by the
Berezinskii-Kosterlitz-Thouless (BKT) theory~\cite{KT-73,B-72}, which
is not associated with any spontaneous symmetry breaking and emergence
of nonvanishing order parameter.  Indeed, 2D fluids of identical
bosons cannot undergo BEC, giving rise to a quasi-long range order at
sufficiently low temperature, characterized by a power-law
large-distance decay of the one-particle correlation function.
Experimental evidences of superfluid transitions in trapped quasi-2D
atomic gases have been reported in
Refs.~\onlinecite{HKCBD-06,KHD-07,HKCRD-08,CRRHP-09,HZGC-10}.  Again,
the trap gives rise to a substantial distortion of the BKT critical
behavior of homogenous systems, see e.g. Ref.~\onlinecite{CR-12}.  The
analysis of the BKT renormalizaton-group flow~\cite{PV-13} shows that
the asymptotic TSS at the BKT transition is characterized by an
apparently trivial trap exponent $\theta=1$, but the asymptotic
behaviors show important multiplicative logarithms. For example, at
$T_c$ the trap correlation length is expected to increase as
$\xi_t\sim l_t (\ln l_t)^\kappa $ with $\kappa=-1$ in the case of
harmonic traps.

\acknowledgements The QMC simulations were performed at the INFN Pisa
GRID DATA center, using also the cluster CSN4.  In total, simulations
took approximately 50 years of CPU time on a single core of a recent
standard commercial processor (most of them devoted to the
simulations in the presence of the trap).

\appendix

\section{Derivation of the trap  exponent $\theta$ }
\label{thetaest}

The {\em trap} exponent $\theta$ generally depends on the universality
class of the transition, on the space dependence of the potential, and
on the way it couples to the particles.
In the case of 3D systems of interacting bosonic particles at the
transition driven by the formation of BEC, the value of $\theta$ can
be inferred by a renormalization-group (RG) analysis of the
perturbation $P_V$ representing the external trapping potential
coupled to the particle density. Let us consider a generic trapping
potential
\begin{equation}
V(r)=v^p r^p
\label{potentialp}
\end{equation}
where $p$ is a positive even interger number (in the case of a
harmonic potential $p=2$), with a trap size defined as $l=J^{1/p}/v$.
We follow the field-theoretical approach of
Refs.~\onlinecite{CV-09,CV-10}, that is we consider the 3D $\Phi^4$
quantum field theory which represents the 3D XY universality class,
see e.g.  Ref.~\onlinecite{ZJ-book},
\begin{equation}
H_{\Phi^4} = \int d^3 x                                        
\left[ |\partial_\mu \psi({\bf x})|^2 + 
r |\psi({\bf x})|^2 + u |\psi({\bf x})|^4\right],
\label{hphi4}
\end{equation} 
where $\psi$ is the complex field associated with the order parameter,
and $r,u$ are coupling constants.  Since the particle density
corresponds to the energy operator $|\psi|^2$, we can write the
perturbation $P_V$ as
\begin{equation}
P_V=\int d^3 x\, V({\bf x}) |\psi({\bf x})|^2.
\label{pertu}
\end{equation}
Introducing the RG dimension $y_v$ of the constant $v$ of the
potential (\ref{potentialp}), we derive the RG relation
\begin{eqnarray}
py_v - p + y_n = 3,\label{rg1}
\end{eqnarray} 
where $y_n=3-1/\nu$ is the RG dimension of the density/energy operator
$|\psi|^2$. We eventually obtain
\begin{equation}
\theta = {1\over y_v}= {p\nu\over 1 + p \nu},
\label{thetap}
\end{equation}
which gives Eq.~(\ref{theta}) in the case of a harmonic potential with
$p=2$.  Note that Eq.~(\ref{thetap}) formally gives $\theta=1$ in the
limit $p\to\infty$.  This is consistent with the fact that the limit
$p\to\infty$ corresponds to a homogenous system in a spherical box of
size $l$, and open boundary conditions. Thus, TSS must reproduce the
standard finite-size scaling of homogenous systems for $p\to\infty$,
where the size $L=l$ behaves as a scaling field of dimension one in
length units.


\begin{thebibliography}{99}
\bibitem{footnoteY} The helicity modulus $\Upsilon$ is related to the
  change of the partition function $Z$ under twists $\phi$ of the
  boundary conditions along one of the lattice directions, i.e.
  $R_\Upsilon\equiv \Upsilon L = - \partial^2 \ln Z(\phi)/\partial
  \phi^2|_{\phi=0}$.
\end{thebibliography}
\end{document}